# Zirconium Carbide as a High-Temperature Benchmark for the Beyond Quasi-Harmonic Method

Christopher M. Stanley,
stanleyc@uindy.edu
University of Indianapolis,
RB Annis School of Engineering
1400 E. Hanna Ave.
Indianapolis, IN 46227

## Abstract

Molar heat capacity at constant pressure, $C_p(T)$, is a central thermodynamic quantity in materials science, but it remains difficult to calculate accurately from first principles when anharmonic vibrational effects become important. In this work, rocksalt ZrC is used as a high-temperature benchmark for the Beyond Quasi-Harmonic (BQH) method, a prepared-supercell first-principles approach that extracts anharmonic vibrational energy directly from density-functional-theory energy evaluations. The calculated BQH molar heat capacity is compared with CALPHAD reference values, a fully anharmonic thermodynamic-integration calculation, a calculated $C_v(T)$ curve from the present work, and a quasi-harmonic calculation. The BQH curve gives a substantial improvement beyond $C_v(T)$ and beyond the quasi-harmonic approximation. Because ZrC is electronically conductive, an electronic heat-capacity correction was also estimated from the density of states at the Fermi level and added to the vibrational BQH result. This correction brings the calculated molar heat capacity into close agreement with the fully anharmonic theoretical benchmark through approximately 1200 K. These results show that the BQH method fully captures important anharmonic phonon contributions missing from ordinary quasi-harmonic calculations.

## 1. Introduction

Heat capacity at constant pressure, $C_p(T)$, is a fundamental thermodynamic quantity for materials science. It is used to obtain temperature-dependent enthalpy and entropy contributions by thermodynamic integration, and these quantities enter Gibbs-energy functions used in phase stability, phase equilibria, and computational thermodynamics [14,15]. Recent work illustrates the breadth of this role: heat-capacity and enthalpy data remain central inputs in modern CALPHAD assessments [20]; heat capacity appears directly as a derivative of thermodynamic potentials in phase-stability calculations and CALPHAD model optimization [21]; and finite-temperature first-principles thermodynamics increasingly relies on free-energy and heat-capacity calculations to construct phase diagrams and evaluate material stability [22]. Accurate heat-capacity data are therefore required not only for thermodynamic tables, but also for CALPHAD assessments, finite-temperature first-principles modeling, phase-diagram construction, and high-temperature materials design.

Zirconium carbide, ZrC, is a useful test material because it is both technologically important and theoretically demanding. It is a member of the ultra-high-temperature ceramic family, where strong metal-nonmetal bonding, high melting temperature, and good high-temperature thermomechanical behavior motivate its use in extreme environments [16]. Recent discussions of ZrC have emphasized hypersonic and aerospace applications, including wing-leading-edge and thermal-protection concepts [17], as well as nuclear-fuel applications where ZrC has been considered as a coating, oxygen getter, or inert matrix material for advanced high-temperature fuels [18]. ZrC is also relevant to the design of ultra-high-temperature ceramic-matrix composites for extreme environments [19]. These applications require accurate thermodynamic data at temperatures where harmonic phonons alone may be insufficient.

The calculation of $C_p(T)$ from first principles is more complicated than the calculation of $C_v(T)$. Harmonic phonon calculations readily provide the constant-volume vibrational heat capacity, but $C_p(T)$ also includes thermal expansion and anharmonic vibrational contributions. The quasi-harmonic approximation (QHA) incorporates part of this physics by evaluating volume-dependent phonon frequencies and minimizing a temperature-dependent Helmholtz free energy. This can be a useful approximation for stiff crystalline solids, but it remains a harmonic phonon theory at each fixed volume. It therefore does not explicitly include non-volume anharmonicity, phonon-phonon interactions, temperature-induced phonon renormalization, or defect-related high-temperature effects.

A more complete route is thermodynamic integration. Duff [4] calculated high-temperature thermodynamic properties of ZrC using two-stage upsampled thermodynamic integration using Langevin dynamics (TU-TILD), including fully anharmonic vibrational contributions and electronic excitations. Subsequent ZrC studies by Mellan [7-9] examined high-temperature point defects, vacancy thermodynamics in substoichiometric $ZrC_x$, and electron-phonon and phonon-phonon transport mechanisms. These works make ZrC a useful benchmark because they provide a high-quality first-principles context for finite-temperature behavior, but the computational setup is substantially more specialized than a standard harmonic or quasi-harmonic calculation.

The BQH method was introduced previously for diamond Si and wurtzite GaN [1], then applied to SiC and Ge [2], and more recently used in a broader heat-capacity optimization study [3]. The method recovers fully anharmonic vibrational energy contributions from prepared supercell configurations without requiring long molecular-dynamics trajectories or a thermodynamic-integration pathway. The present work applies BQH to rock salt ZrC in order to test whether this simpler prepared-supercell method provides a meaningful improvement over QHA for a refractory, electronically conductive carbide.

## 2. Theory and Computational Method

### 2.1. Beyond Quasi-Harmonic Method

The BQH method is based on prepared supercells and has been described in detail previously [1-3,12, 23-26]. A supercell is prepared in a linear combination of harmonic normal modes, with mode amplitudes and phases corresponding to a selected temperature distribution. This preparation gives the system a known harmonic vibrational energy. An example of this process using Silicon Carbide as well as the Prepcell™ code and instructions for its use can be found here in ref. [23]. The displaced configuration is then evaluated using density functional theory. Since the Kohn-Sham energy contains the response of the electronic ground state to the full displaced atomic configuration, while the preparation procedure accounts only for harmonic vibrational energy, the difference between the DFT energy and the harmonic energy is then the anharmonic energy contribution [1,2]. Figure 1 illustrates this idea qualitatively for a single normal-mode coordinate. But of course, for the system level configuration, all normal vibrational modes are excited simultaneously, so figure 1 is for illustrative purposes only.

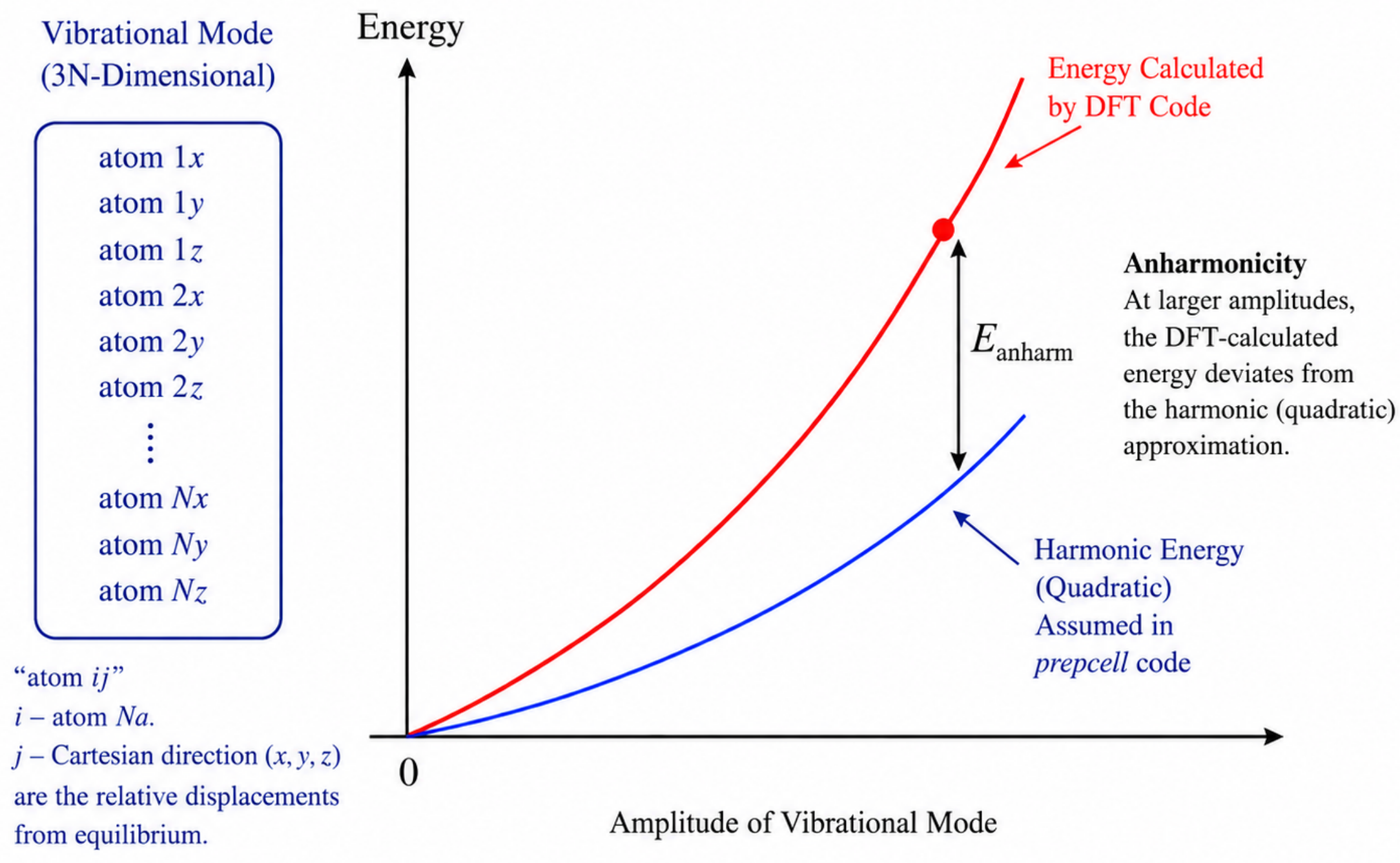

Figure 1. Qualitative schematic of the BQH/prepared-supercell energy extraction. The prepared-supercell code assigns a harmonic quadratic energy, based on the desired temperature, to a displacement along a normal-mode coordinate/Amplitude, while the DFT single-point calculation evaluates the true value of the system energy. The excess energy is identified as the anharmonic vibrational contribution, $E_{\mathrm{anharm}}$. As discussed in ref. [1] the desired temperature is mapped to an amplitude for each mode, yielding a specific configuration, that is given to SIESTA which outputs the true system energy.

Conceptually, the total vibrational energy is written as

$$E_{\mathrm{total}} = E_{\mathrm{harm}} + E_{\mathrm{anharm}}. \qquad (1)$$

The harmonic phonon contribution was calculated from the phonon density of states, $g(\omega)$, using the Helmholtz free energy expression used in the original BQH work [1]:

$$F(T) = k_B T \int_0^\infty \ln\left[2\sinh\left(\frac{\hbar\omega}{2k_B T}\right)\right] g(\omega)\, d\omega. \qquad (2)$$

The corresponding harmonic constant-volume heat capacity is then

$$C_v(T) = -T\frac{\partial^2 F}{\partial T^2}. \qquad (3)$$

Once $E_{\mathrm{anharm}}(T)$ has been calculated over a set of temperatures and fitted with a smooth function (see Figure 3 below), the vibrational BQH heat capacity is obtained from

$$C_{p,\mathrm{BQH}}(T) = C_v(T) + \frac{dE_{\mathrm{anharm}}}{dT}. \qquad (4)$$

Here $C_v(T)$ is supplied by the harmonic phonon calculation, while the derivative of the fitted anharmonic energy supplies the anharmonic correction. In contrast, QHA obtains its correction primarily from the volume dependence of harmonic phonons. The key test in the present work is therefore whether the prepared-supercell BQH correction captures heat-capacity contributions that are absent or underestimated in QHA, and how well at high temperatures.

### 2.2. Hot-Block Construction

In the present implementation, a hot block is defined within the supercell, but the entire cell is affected by the preparation. The hot block is excited by stretching the relevant local vibrational amplitudes, while the surrounding portion of the supercell responds elastically as the hot block expands into it. Thus, the rest of the supercell is colder only in the harmonic bookkeeping sense; electronically and mechanically, it is part of the same displaced DFT configuration and can be compressed by the expansion of the hot region. Periodic boundary conditions mean that each nominally colder region is also sandwiched between repeated images of the same locally heated region. In other words the system is a cube broken into 8 sections, which one of them (the hot block) expanding via increased normal mode amplitudes.

Heating the entire supercell uniformly would force the calculation to be a strictly constant volume practically by definition as a direct consequence of the periodic boundary conditions. When this simulation is performed, no addition energy above the harmonic approximation is ever found, thus the exercise is not very useful. The hot-block construction avoids this problem by allowing a locally excited region to expand into surrounding material, while also allowing that material to be compressed. For the present ZrC calculation, the cubic 64-atom rocksalt supercell used an 8-atom hot block, corresponding to one eighth of the supercell. This is the same fractional hot-block size used for cubic crystals in the earlier BQH works [1 - 3].

### 2.3. Density Functional Theory Settings

All electronic-structure calculations were performed with SIESTA version 4.1.5 [11,13] using norm-conserving pseudopotentials and numerical atomic orbitals. The model system was a 64-atom rock salt ZrC supercell containing 32 C atoms and 32 Zr atoms. The lattice constant was optimized to be 4.7081465726 Å, and the SIESTA lattice vectors were given as $(2,0,0)$, $(0,2,0)$, and $(0,0,2)$ in units of the lattice constant, corresponding to a cubic $2 \times 2 \times 2$ conventional supercell. The exchange-correlation functional was PBE-GGA, and the mesh cutoff was 350 Ry. The carbon atoms were treated with a double-zeta basis with a PAO energy shift of 0.02eV. Zr was treated with an explicitly specified basis including $4s$, $4p$, $4d$, $5s$, and $5p$ orbitals. The Zr basis contains relatively diffuse cutoff radii, including several radii of approximately 9 Bohr. Full settings used in the calculation can be found in the github repo linked in ref. [27].

### 2.4. Electronic Heat-Capacity Correction

Because ZrC has no band gap, the vibrational heat capacity is not the only possible finite-temperature contribution. In a metal, thermal occupation of electronic states near the Fermi level gives an electronic heat capacity that, to leading order, is proportional to temperature. This contribution was estimated from a separate electronic density-of-states calculation using the same 64-atom ZrC supercell, with a denser (6x6x6) shifted Monkhorst-Pack grid for Brillouin-zone sampling.

The leading electronic heat-capacity coefficient is

$$\gamma = \frac{\pi^2}{3} k_B^2 D(E_F), \qquad (5)$$

where $D(E_F)$ is the electronic density of states at the Fermi level. In molar units, if $D(E_F)$ is expressed in states/eV per ZrC formula unit, this becomes

$$\gamma = 0.002357D(E_F) \quad \mathrm{J\,mol^{-1}\,K^{-2}}. \qquad (6)$$

The electronic contribution was then approximated as

$$C_{V,\mathrm{el}}(T) \approx \gamma T. \qquad (7)$$

Finally, the electronically corrected BQH result was calculated as

$$C_{p,\mathrm{BQH+el}}(T) = C_{p,\mathrm{BQH}}(T) + C_{V,\mathrm{el}}(T). \qquad (8)$$

The DOS calculation gave $D(E_F) \approx 6.54$ states/eV for the 64-atom supercell. Since the cell contains 32 ZrC formula units, this corresponds to $D(E_F) \approx 0.204$ states/eV per formula unit and $\gamma \approx 4.82 \times 10^{-4}\ \mathrm{J\,mol^{-1}\,K^{-2}}$. The resulting correction is approximately 0.58 $\mathrm{J\,mol^{-1}\,K^{-1}}$ at 1200 K and 0.72 $\mathrm{J\,mol^{-1}\,K^{-1}}$ at 1500 K. This is a small correction relative to the total molar heat capacity, but it is large enough to matter when comparing high-quality theoretical curves.

This treatment is a first-order electronic correction rather than a full finite-temperature electronic free-energy calculation. It assumes that the DOS is sufficiently smooth near $E_F$ for the Sommerfeld form to be useful over the temperature range considered. A more rigorous treatment would integrate the finite-temperature electronic energy and entropy using the full DOS near $E_F$. The present correction is nevertheless sufficient to estimate the leading metallic electronic contribution and to determine whether electronic excitations materially affect the comparison to reference calculations that include them.

### 2.5. Reference Data and Comparison Calculations

Four comparison datasets are used in the heat-capacity figures: the thermochemical CALPHAD data [5], the TU-TILD results [4], the quasi-harmonic calculation [6], and the present calculated $C_v(T)$, the latter being included mostly for reference purposes. Duff [4] is the most direct theoretical benchmark because it includes fully anharmonic vibrational contributions and first-principles electronic excitations.

## 3. Results and Discussion

### 3.1. Heat Capacity of ZrC

Figure 2 shows the calculated molar heat capacity of ZrC, reported in $\mathrm{J\,mol^{-1}\,K^{-1}}$. Panel (a) gives the full plotted temperature range, while panel (b) expands the region where the differences among $C_v(T)$, QHA, BQH, and the comparison datasets are most visible.

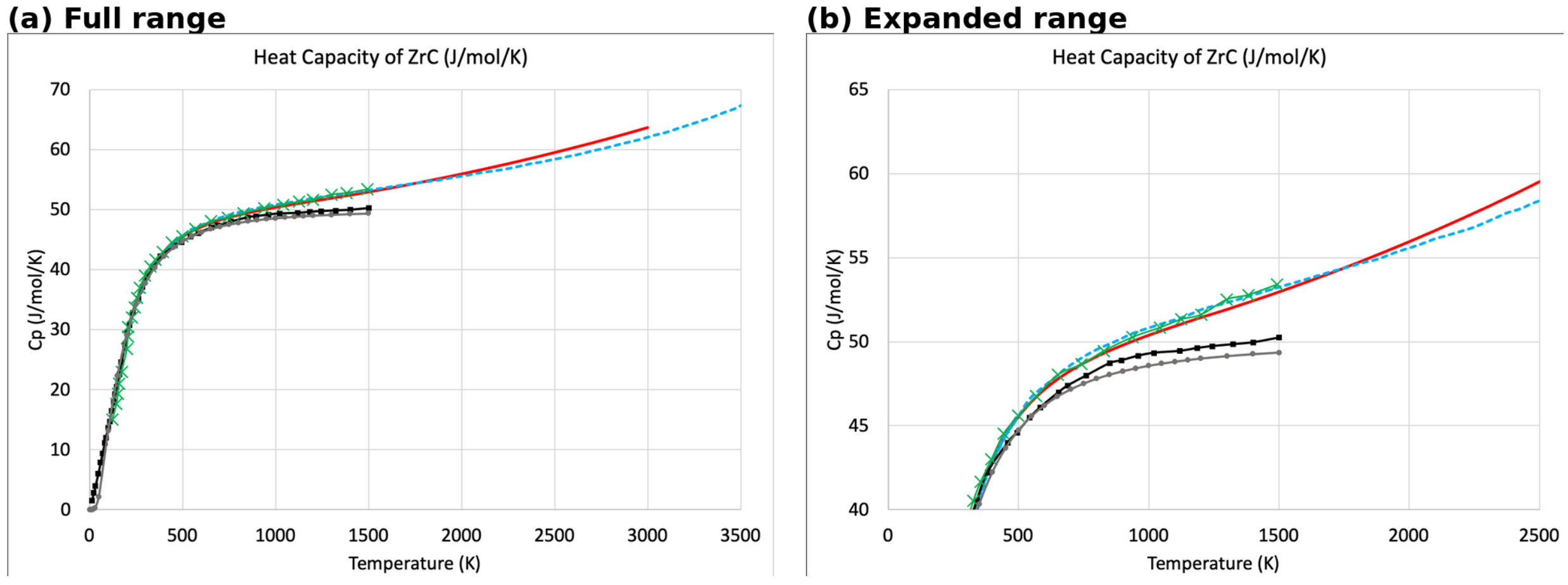

Figure 2. Molar heat capacity of ZrC. (a) Full plotted temperature range. (b) Expanded comparison. Duff [4] is the dashed blue line; the BQH method (this work), is the solid red line; the green x symbols are the CALPHAD data [5]; the solid gray line with filled circles is the present $C_v(T)$ calculation; and the black solid line with black squares is the quasi-harmonic data from Yang [6]. The red BQH curve is extrapolated beyond the max temperature (1600K) used to fit $E_{\text{anharm}}(T)$, and thus is shown here only for the sake of discussion.

An important feature of Fig. 2b is the separation between the QHA curve and the BQH/TU-TILD/CALPHAD curves. The QHA result lies much closer to the present $C_v(T)$ calculation than to either the first principles curves or experimental data. It therefore provides only a modest improvement over the harmonic, constant-volume calculation. This indicates that, for the ZrC data considered here, volume-dependent harmonic phonons alone do not fully describe the molar heat capacity. This interpretation is consistent with Yang [6], who connected the increasing high-temperature discrepancy in the QHA result with missing anharmonic effects.

By contrast, the BQH method gives a much larger correction beyond $C_v(T)$ because it samples the displaced DFT energy surface directly and directly accounts for anharmonic terms not represented in a fixed harmonic phonon calculation. This is the central motivation for applying BQH: the method is worth using only if it captures physics that QHA misses. In ZrC, as previously observed for GaN [1] and Ge [2], the BQH method consistently provides a substantial improvement over QHA.

Figure 3 shows the fitted anharmonic energy used to generate $E_{\text{anharm}}(T)$ and by extension $\frac{dE_{\text{anharm}}}{dT}$ and of course $C_{p,\text{BQH}}(T)$ itself. The fit itself is $E_{\text{anharm}}(T)$, and the derivative provides the second term in Eq. (4).

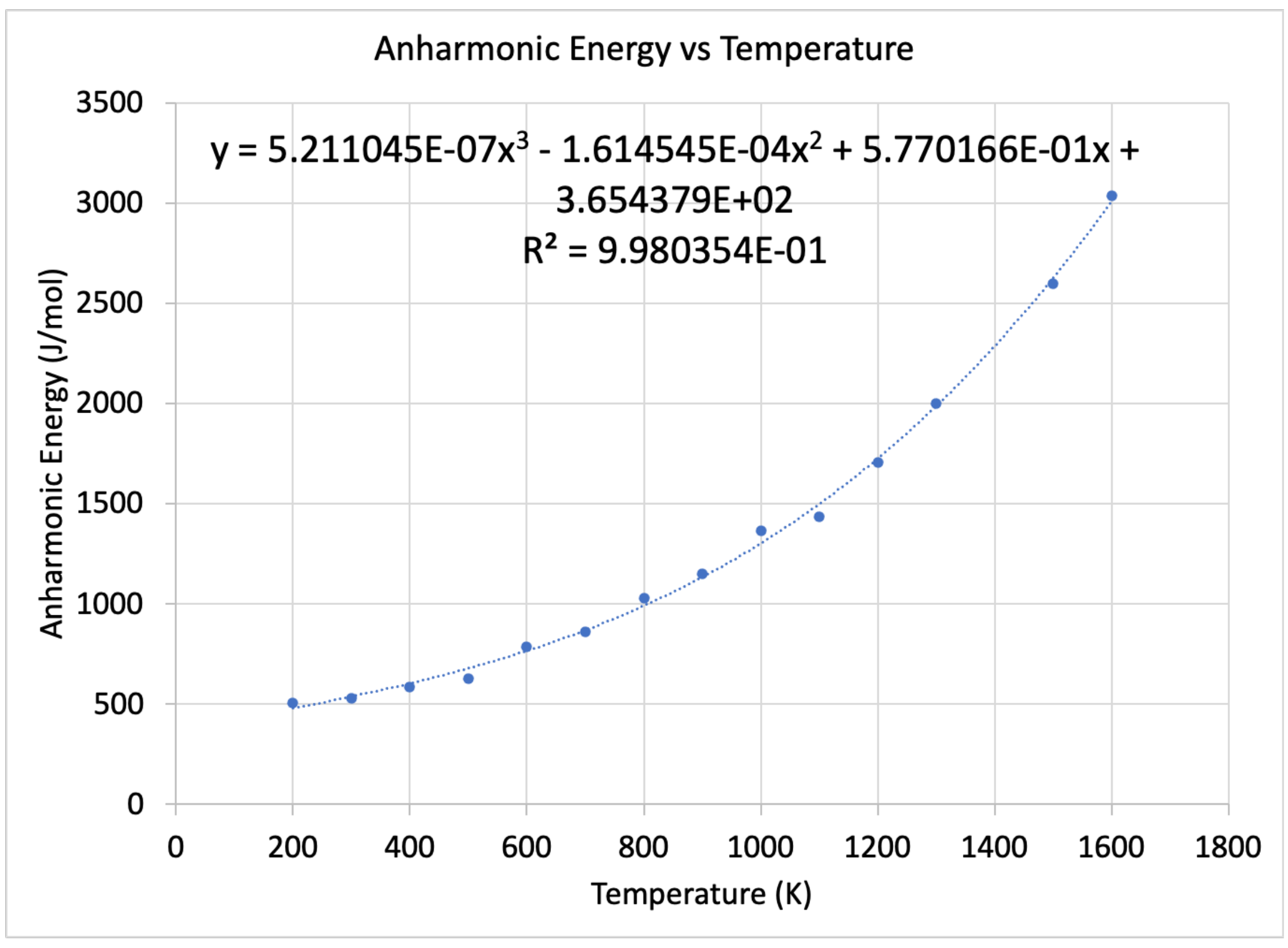

Figure 3. Anharmonic energy extracted from the prepared-supercell and DFT/SIESTA calculations as a function of temperature. The plotted polynomial fit is the function $E_{\text{anharm}}(T)$ which is then differentiated with respect to temperature to obtain the BQH anharmonic heat-capacity in Eq. (4).

### 3.2. Effect of the Electronic Correction

Figure 4 shows the effect of adding the electronic correction estimated from the DOS at the Fermi level. The solid purple line is $C_{p,\mathrm{BQH+el}}(T)$, calculated from Eq. (8). The correction is numerically small, but it systematically raises the BQH curve by the amount expected for a metallic electronic heat-capacity contribution.

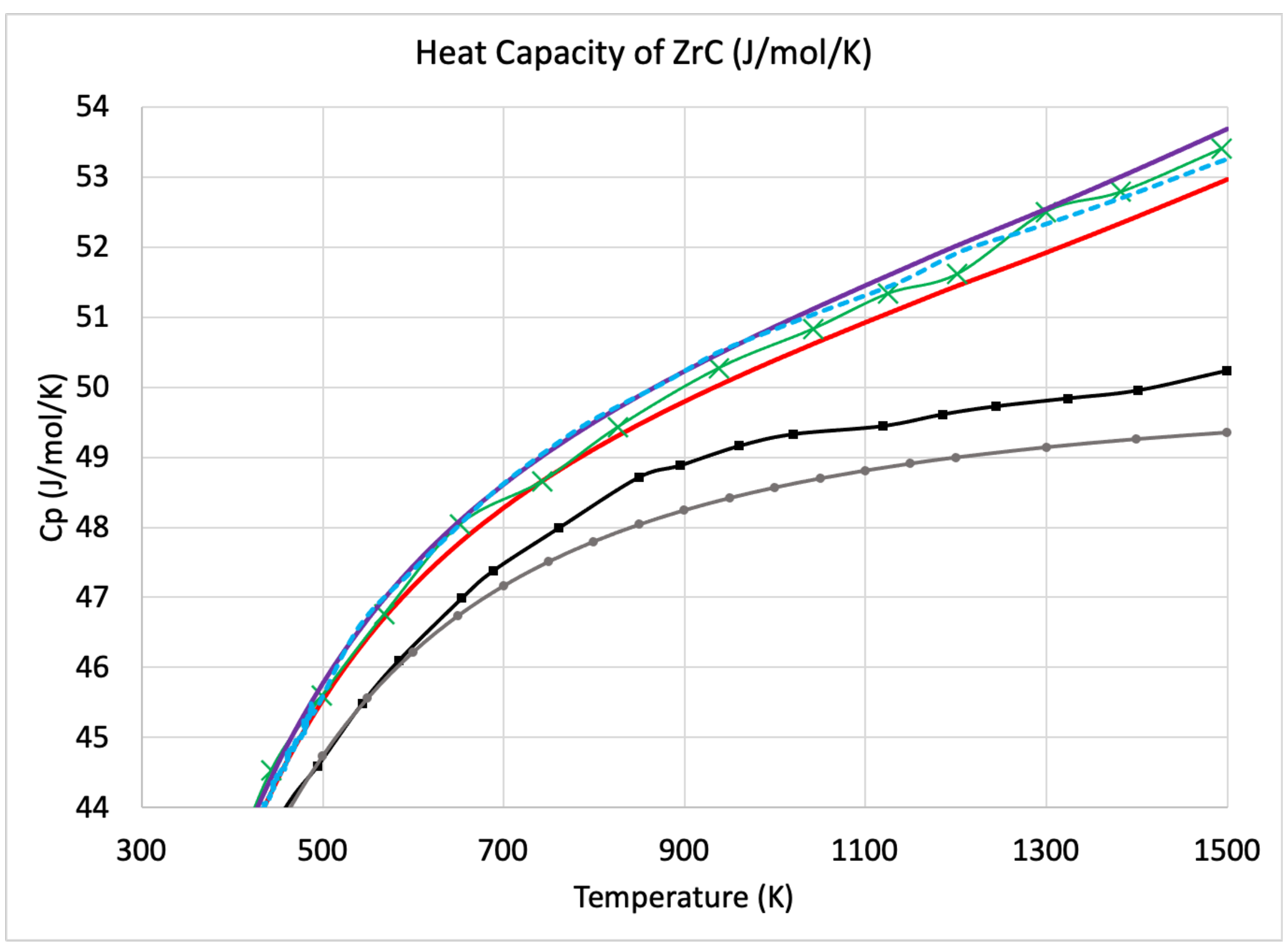

Figure 4. Expanded molar heat-capacity comparison for ZrC including the electronic correction. The solid purple line is the electronically corrected BQH result, $C_{p,\mathrm{BQH+el}}(T)$. The solid red line is the uncorrected BQH vibrational result. The other data have the same meanings as in Fig. 2.

The electronically corrected curve in Figure 4 is nearly identical to the TU-TILD results up until ~1000K, with close agreement found to be quite good for the rest of the graph. This is important because Duff [4] includes electronic excitations, while the uncorrected BQH calculation is based on phonon contributions only. The data begin to deviate consistently around 1300K. It's possible (likely?) that the deviation at higher temperature is related to the limits of the simple correction that we used (see equations 5 – 8), especially if the electronic DOS is not flat and/or smooth over the thermal energy window around the Fermi level.

This result also clarifies the role and limits of DFT in the BQH calculation. The DFT single-point energy of a distorted supercell describes the lowest-energy electronic response to a distorted nuclear geometry. It therefore captures orbital deformation, bond stretching, and bond compression on the ground-state electron energy surface. And in a sense, can capture "excited" electron states. But it is fundamentally limited by the fact that DFT is by its nature an energy minimization (hence the F, for "functional"). Thus it does not automatically include the finite-temperature occupation of electronic states above the Fermi level. But as we can this this effect is much smaller than one might immediately expect exactly because distorted electron orbitals that it can describe are, in some sense excited. That said ZrC is, at the end of the day, electronically conductive, and so while the leading electronic heat-capacity term is small, it is clearly visible, and it grows with increasing temperature.

## 4. Conclusions

ZrC has been used as a benchmark for the Beyond Quasi-Harmonic method. The present comparison shows that the BQH calculation gives a significant improvement over the quasi-harmonic approximation, exactly because it takes into account all anharmonic interactions—both phonon-phonon interactions in addition to effects from volume changes. Thus it is useful especially for materials where the quasi harmonic method is insufficient, like we've shown that it is here with rock salt ZrC; and like we showed previously with Ge [2]. In addition, even when QHA does a decent job the BQH method still outperforms it [3], while still being far simpler than the TU-TILD method. At this point the BQH method has been verified on a wide variety of materials crystal structures including diamond structure Si[1] and Ge[2], Zincblend SiC[2], wurtzite GaN [1], AlN [3], BN [3], and now rock salt ZrC (this work).

Future work will likely focus on including a full, finite-temperature, electron free-energy correction rather than only the Sommerfeld approximation used here. Furthermore, extending the temperature range of the method will also likely be a priority, since the current method as it stand now only produces results to ~1600K. In addition, we will likely wish to further verify the BQH method on different material types such as amorphous systems like a-$SiO_2$ or van der Waals materials like graphite, as well as nanostructures and biomolecules. These extensions might even allow for direct calculations on 2D materials like carbon nanotubes and graphene.

## Acknowledgements:

This work used **Stampede3** at **Texas Advanced Computing Center** through allocation **PHY210056** from the Advanced Cyberinfrastructure Coordination Ecosystem: Services & Support (ACCESS) program, which is supported by U.S. National Science Foundation grants #2138259, #2138286, #2138307, #2137603, and #2138296.

## References

1. Stanley, C. M. Specific heat at constant pressure from first principles: contributions from fully anharmonic vibrations. *Materials Research Express* **6**, 125924 (2019). DOI: 10.1088/2053-1591/ab7157.

2. Stanley, C. M. Vibrational enthalpies of solid crystalline materials. *Solids* **3**, 319-326 (2022). DOI: 10.3390/solids3020023.

3. Vassiliev, V. P.; Stanley, C. M. Optimization of heat capacities of wurtzite phases as a single system and thermodynamic properties of nihonium nitride. *Calphad* **89**, 102824 (2025). DOI: 10.1016/j.calphad.2025.102824.

4. Duff, A. I.; Davey, T.; Korbmacher, D.; Glensk, A.; Grabowski, B.; Neugebauer, J.; Finnis, M. W. Improved method of calculating ab initio high-temperature thermodynamic properties with application to ZrC. *Physical Review B* **91**, 214311 (2015). DOI: 10.1103/PhysRevB.91.214311.

5. Fernandez Guillermet, A. Analysis of thermochemical properties and phase stability in the zirconium-carbon system. *Journal of Alloys and Compounds* **217**, 69-89 (1995). DOI: 10.1016/0925-8388(94)01310-E.

6. Yang, X.-Y.; Lu, Y.; Zheng, F.-W.; Zhang, P. Mechanical, electronic, and thermodynamic properties of zirconium carbide from first-principles calculations. *Chinese Physics B* **24**, 116301 (2015). DOI: 10.1088/1674-1056/24/11/116301.

7. Mellan, T. A.; Duff, A. I.; Finnis, M. W. Spontaneous Frenkel pair formation in zirconium carbide. *Physical Review B* **98**, 174116 (2018). DOI: 10.1103/PhysRevB.98.174116.

8. Mellan, T. A.; Duff, A. I.; Grabowski, B.; Finnis, M. W. Fast anharmonic free energy method with an application to vacancies in ZrC. *Physical Review B* **100**, 024303 (2019). DOI: 10.1103/PhysRevB.100.024303.

9. Mellan, T. A.; Aziz, A.; Xia, Y.; Grau-Crespo, R.; Duff, A. I. Electron and phonon interactions and transport in the ultrahigh-temperature ceramic ZrC. *Physical Review B* **99**, 094310 (2019). DOI: 10.1103/PhysRevB.99.094310.

10. Tiwari, J.; Feng, T. Intrinsic thermal conductivity of ZrC from low to ultrahigh temperatures. *Physical Review Materials* **7**, 065001 (2023). DOI: 10.1103/PhysRevMaterials.7.065001.

11. Soler, J. M.; Artacho, E.; Gale, J. D.; Garcia, A.; Junquera, J.; Ordejon, P.; Sanchez-Portal, D. The SIESTA method for ab initio order-N materials simulation. *Journal of Physics: Condensed Matter* **14**, 2745-2779 (2002).

12. Gibbons, T. M.; Bebek, M. B.; Kang, B.; Stanley, C. M.; Estreicher, S. K. Phonon-phonon interactions: First-principles theory. *Journal of Applied Physics* **118**, 085103 (2015). DOI: 10.1063/1.4929487.

13. Garcia, A. et al. Siesta: Recent developments and applications. *Journal of Chemical Physics* **152**, 204108 (2020). DOI: 10.1063/5.0005077.

14. Jofre, J.; Gheribi, A. E.; Harvey, J.-P. Development of a flexible quasi-harmonic-based approach for fast generation of self-consistent thermodynamic properties used in computational thermochemistry. *Calphad* **83**, 102624 (2023). DOI: 10.1016/j.calphad.2023.102624.

15. Jung, J. H.; Srinivasan, P.; Forslund, A.; Grabowski, B. High-accuracy thermodynamic properties to the melting point from ab initio calculations aided by machine-learning potentials. *npj Computational Materials* **9**, 3 (2023). DOI: 10.1038/s41524-022-00956-8.

16. Wyatt, B. C.; Nemani, S. K.; Hilmas, G. E.; Opila, E. J.; Anasori, B. Ultra-high temperature ceramics for extreme environments. *Nature Reviews Materials* **9**, 773-789 (2024). DOI: 10.1038/s41578-023-00619-0.

17. Peterson, G. R.; Carr, R. E.; Marinero, E. E. Zirconium carbide for hypersonic applications, opportunities and challenges. *Materials* **16**, 6158 (2023). DOI: 10.3390/ma16186158.

18. Kato, Y.; Vasudevamurthy, G.; Nozawa, T.; Rana, D.; Farnan, I.; Snead, L. Properties of zirconium carbide for nuclear fuel applications. In *Comprehensive Nuclear Materials*, 2nd ed.; Elsevier, 419-456 (2020).

19. Zhou, Y. From properties of zirconium carbide to the reverse design of ultra-high temperature CMCs. *Journal of the American Ceramic Society* **107**, 7023-7037 (2024). DOI: 10.1111/jace.20016.

20. Ali, M. L.; Ramos, S. B.; Fernandez Guillermet, A. J. Heat capacity and enthalpy of palladium: A critical analysis of experimental information. *Calphad* 84, 102670 (2024). DOI: 10.1016/j.calphad.2024.102670.

21. Kunselman, C.; Bocklund, B. J.; van de Walle, A.; Otis, R. D.; Arroyave, R. Analytically differentiable metrics for phase stability. *Calphad* 86, 102705 (2024). DOI: 10.1016/j.calphad.2024.102705.

22. Monacelli, L.; Marzari, N. First-principles thermodynamics of CsSnI3. *Chemistry of Materials* 35, 1702-1709 (2023). DOI: 10.1021/acs.chemmater.2c03475.

23. The github repo of the most recent prepcell release: https://github.com/ProfStanley/Prepcell/releases/tag/v2.0.0

24. Bebek, M., Stanley, C., Gibbons, T. et al. Temperature dependence of phonon-defect interactions: phonon scattering vs. phonon trapping. Sci Rep 6, 32150 (2016). https://doi.org/10.1038/srep32150

25. Stanley, C.M., Rader, B.K., Laster, B.H.D., Servati, M. and Estreicher, S.K. (2021), The Role of Interface Vibrational Modes in Thermal Boundary Resistance. Phys. Status Solidi A, 218: 2100111. https://doi.org/10.1002/pssa.202100111

26. Stanley, C.M. and Estreicher, S.K. (2019), Phonon Dynamics at an Oxide Layer in Silicon: Heat Flow and Kapitza Resistance. Phys. Status Solidi A, 216: 1800428. https://doi.org/10.1002/pssa.201800428

27. ZrC calculations and code repo: https://github.com/ProfStanley/rock-salt-ZrC-Cp-calc/releases/tag/ZrC64_paper_v1.0